\documentclass[12pt,a4paper]{article}
\usepackage[utf8]{inputenc}
\usepackage[english]{babel}
\usepackage{amsmath,amssymb}
\usepackage{bm}
\usepackage{graphicx}
\usepackage{array}
\usepackage{geometry}
\usepackage{hyperref}
\hypersetup{colorlinks=true, linkcolor=blue, citecolor=blue, urlcolor=blue}

\makeatletter
\renewcommand{\maketitle}{\par
  \begingroup
    \renewcommand\thefootnote{\@fnsymbol\c@footnote}%
    \def\@makefnmark{\rlap{\@textsuperscript{\normalfont\@thefnmark}}}%
    \long\def\@makefntext##1{\parindent 1em\noindent
            \hb@xt@1.8em{\hss\@textsuperscript{\normalfont\@thefnmark}}##1}%
    \newpage
    \global\@topnum\z@
    \@maketitle
    \thispagestyle{plain}\@thanks
  \endgroup
  \setcounter{footnote}{0}%
  \global\let\thanks\relax
  \global\let\maketitle\relax
  \global\let\@maketitle\relax
  \global\let\@thanks\@empty
  \global\let\@author\@empty
  \global\let\@title\@empty
  \global\let\title\relax
  \global\let\author\relax
  \global\let\date\relax
  \global\let\and\relax
}
\def\@maketitle{%
  \newpage
  \null
  \vskip -2em
  \begin{center}%
  \let \footnote \thanks
    {\LARGE \@title \par}%
    \vskip 1.5em%
    {\large
      \lineskip .5em%
      \begin{tabular}[t]{c}%
        \@author
      \end{tabular}\par}%
    \vskip 1em%
    {\large \@date}%
  \end{center}%
  \par
  \vskip 1.5em}
\makeatother

\title{\Large\textbf{Control Algorithms for Quadcopter Motion in Dynamic Positioning Mode}}
\author{
    \textbf{S. A. Kim}$^{1}$, \textbf{A. A. Pyrkin}$^{1}$, \textbf{O. I. Borisov}$^{1}$ \\[2mm]
    $^{1}$ITMO University, Saint Petersburg, Russia \\
    \texttt{skim@itmo.ru}
}
\date{}  

\begin{document}

\maketitle

\begin{abstract}
A complete model of quadcopter motion for the task of dynamic positioning at a specified point is derived. Based on this model, two control algorithms are proposed. The first one generalizes previously obtained results to the case of a varying yaw angle. The second control algorithm addresses the above problem using a simplified regulator tuning methodology.
\end{abstract}

\textbf{Acknowledgements:} This work was supported by the Grant of the President of the Russian Federation No. MD-3574.2022.4 and the Ministry of Science and Higher Education of the Russian Federation (state assignment passport No. 2019-0898). \\[1mm]

\section{Introduction}

The nonlinear dynamics, multi-loop structure, and the presence of uncertainties make the challenge of controlling quadcopter as relevant as ever. The classical approach involves model linearization to simplify regulator synthesis; however, such an approach is rather crude and, due to model inaccuracies, does not always provide satisfactory transient performance. The related work, for example, \cite{1,2} suggests solutions that consider the nonlinear motion model, however, they use an incomplete motion model that does not account for the yaw angle dynamics. Moreover, the proposed control algorithms assume a non-obvious tuning methodology, where the regulator parameter values must be chosen to satisfy the conditions related to the proofs of statements.

This work addresses the identified shortcomings, resulting in a complete model of quadcopter motion for the dynamic point positioning problem, based on which two control algorithms are proposed: 1) generalizing the result of \cite{1} to the case of a varying yaw angle; 2) suggesting a dynamic control algorithm with a simplified regulator tuning methodology.

\section{Quadcopter Motion Model in Nonstationary Normal Form}

The spatial motion model has the form:
\begin{align}
\begin{bmatrix} \dot{x} \\ \dot{y} \\ \dot{z} \end{bmatrix}
&= \begin{bmatrix} v_x \\ v_y \\ v_z \end{bmatrix}, \\[2mm]
\begin{bmatrix} \ddot{x} \\ \ddot{y} \\ \ddot{z} \end{bmatrix}
&= - \begin{bmatrix} a_x & 0 & 0 \\ 0 & a_y & 0 \\ 0 & 0 & a_z \end{bmatrix}
   \begin{bmatrix} v_x \\ v_y \\ v_z \end{bmatrix}
   + \frac{1}{m}
   \begin{bmatrix}
     \cos\phi\sin\theta\cos\psi + \sin\phi\sin\psi \\
     \sin\phi\sin\theta\cos\psi - \cos\phi\sin\psi \\
     \cos\theta\cos\psi
   \end{bmatrix}
   \sum_{i=1}^{4} F_i
   - \begin{bmatrix} 0 \\ 0 \\ g \end{bmatrix}, \\[2mm]
\begin{bmatrix} \ddot{\psi} \\ \ddot{\theta} \\ \ddot{\phi} \end{bmatrix}
&= -
   \begin{bmatrix} a_{\psi} & 0 & 0 \\ 0 & a_{\theta} & 0 \\ 0 & 0 & a_{\phi} \end{bmatrix}
   \begin{bmatrix} \dot{\psi} \\ \dot{\theta} \\ \dot{\phi} \end{bmatrix}
   + \frac{C}{\sqrt{J_{\psi}}}
   \begin{bmatrix}
     -1 & 1 & 1 & -1 \\
     -1 & -1 & 1 & 1 \\
     1 & -1 & 1 & -1
   \end{bmatrix}
   \begin{bmatrix} F_1 \\ F_2 \\ F_3 \\ F_4 \end{bmatrix},
\end{align}
where \(x, y, z\) are the Cartesian coordinates of the center of mass; \(\phi, \theta, \psi\) are the yaw, pitch, and roll angles; \(g = 9.81\) m/s\(^2\) is the gravitational acceleration; \(m\) is the mass; \(F_i\), \(i = 1\dots 4\) are the rotor thrust forces; \(\ell\) is the distance between the center of gravity and the rotors; \(J_{\psi}, J_{\theta}, J_{\phi}\) are the moments of inertia; \(C\) is the proportionality coefficient; the parameters \(a\) with the corresponding indices in each dynamic channel denote viscous friction coefficients. Since the forces \(F_i\) are sufficiently small in the normal quadcopter operating mode characterized by low velocities, this component of the dynamics equation will be neglected hereafter. Noteworthily, including an additional component would only make the expressions more cumbersome without affecting the key result.

Let us introduce auxiliary variables:
\[
\begin{bmatrix}
\frac{1}{m} & 0 & 0 \\
0 & \frac{C}{J_{\phi}} & 0 \\
0 & 0 & \frac{\ell}{J_{\psi}}
\end{bmatrix}
\begin{bmatrix}
1 & 1 & 1 & 1 \\
1 & -1 & 1 & -1 \\
-1 & 1 & 1 & -1 \\
-1 & -1 & 1 & 1
\end{bmatrix}
\begin{bmatrix}
F_1 \\ F_2 \\ F_3 \\ F_4
\end{bmatrix}
=
\begin{bmatrix}
g \\ 0 \\ 0 \\ 0
\end{bmatrix}
\]
and rewrite the complete dynamic model of the robot's spatial motion in a more compact form:
\begin{align}
\begin{bmatrix} \dot{x} \\ \dot{y} \\ \dot{z} \end{bmatrix}
&= \begin{bmatrix} v_x \\ v_y \\ v_z \end{bmatrix}, \label{eq:kin}\\
\begin{bmatrix} \dot{v}_x \\ \dot{v}_y \\ \dot{v}_z \end{bmatrix}
&= \begin{bmatrix}
     \cos\phi\sin\theta\cos\psi + \sin\phi\sin\psi \\
     \sin\phi\sin\theta\cos\psi - \cos\phi\sin\psi \\
     \cos\theta\cos\psi
   \end{bmatrix}
   \begin{bmatrix} u_1 + g \\ u_2 \end{bmatrix}
   - \begin{bmatrix} 0 \\ 0 \\ g \end{bmatrix}, \label{eq:dyn}
\end{align}
where the control inputs \(u_1, u_2\) are defined via the thrust forces.

Consider the following statement.

\textbf{Statement 1.}
Model \eqref{eq:kin}–\eqref{eq:dyn} can be represented as a cascade of systems in the normal form with nonstationary matrix coefficients in the integrator chain:
\begin{align}
\dot{\xi}_1 &= \xi_2, \\
\dot{\xi}_2 &= q_1(\xi) + b_1(\xi) \begin{bmatrix} u_1 \\ u_2 \end{bmatrix}, \\
\dot{\xi}_3 &= \xi_4, \\
\dot{\xi}_4 &= \beta(t)\xi_5, \\
\dot{\xi}_5 &= \xi_6, \\
\dot{\xi}_6 &= q_2(\xi) + b_{21}(\xi)u_2 + b_{22}(\xi) \begin{bmatrix} u_3 \\ u_4 \end{bmatrix},
\end{align}
where \(\xi = \operatorname{col}(\xi_1, \ldots, \xi_6)\) is the state vector in the new basis, the variable matrices of appropriate dimensions are \(q_1, b_1, q_2, b_{21}, b_{22}\); \(\beta(t)\) is a positive coefficient satisfying
\begin{equation}\label{eq:beta}
0 < \beta_{\min} \le \beta(t) \le \beta_{\max}.
\end{equation}

We write the quadcopter motion model in deviations from the desired position and orientation and present the solution to the problem of stabilizing the quadcopter at a given point in space in the absence of external disturbances. The regulated variables in this case are the linear spatial coordinates and the yaw angle:
\[
Y = \operatorname{col}(z, \phi, x, y) = \operatorname{col}(\xi_1, \xi_3).
\]

The desired position of the quadcopter in space is described by the vector of constant values
\[
Y^* = \operatorname{col}\left( z^*, \phi^*, x^*, y^* \right)
\]
(the asterisk denotes the desired value of the corresponding variable), as well as zero roll and pitch angles:
\[
\psi^* = 0 \quad \text{and} \quad \theta^* = 0.
\]

Let us introduce the stabilization error vector:
\begin{align*}
\tilde{\xi}_1 &=
\begin{bmatrix} \tilde{\xi}_{11} \\ \tilde{\xi}_{12} \end{bmatrix}
= \begin{bmatrix} z - z^* \\ \phi - \phi^* \end{bmatrix}
= \xi_1 - \begin{bmatrix} z^* \\ \phi^* \end{bmatrix}, \quad
\tilde{\xi}_2 = \xi_2,\\
\tilde{\xi}_3 &=
\begin{bmatrix} \tilde{\xi}_{31} \\ \tilde{\xi}_{32} \end{bmatrix}
= \begin{bmatrix} x - x^* \\ y - y^* \end{bmatrix}
= \xi_3 - \begin{bmatrix} x^* \\ y^* \end{bmatrix}, \quad
\tilde{\xi}_4 = \xi_4,\\
\tilde{\xi}_5 &= \xi_5, \quad \tilde{\xi}_6 = \xi_6
\end{align*}
and, using Statement 1, find the dynamic model of quadcopter motion in deviations from the desired position and orientation:
\begin{align}
\dot{\tilde{\xi}}_1 &= \tilde{\xi}_2, \\
\dot{\tilde{\xi}}_2 &= q_1(\psi, \theta) + b_1(\psi, \theta) \begin{bmatrix} u_1 \\ u_2 \end{bmatrix}, \\
\dot{\tilde{\xi}}_3 &= \tilde{\xi}_4, \\
\dot{\tilde{\xi}}_4 &= \beta(t)\tilde{\xi}_5, \\
\dot{\tilde{\xi}}_5 &= \tilde{\xi}_6, \\
\dot{\tilde{\xi}}_6 &= q_2(\phi, \psi, \theta, \dot{\phi}, \dot{\psi}, \dot{\theta}) + b_{21}(\phi, \psi, \theta)u_2 + b_{22}(\phi, \psi, \theta) \begin{bmatrix} u_3 \\ u_4 \end{bmatrix}.
\end{align}

The control synthesis objective is to choose regulators \(u_1, u_2, u_3, u_4\) such that the zero equilibrium \(\tilde{\xi} = \operatorname{col}(\tilde{\xi}_1, \tilde{\xi}_2, \tilde{\xi}_3, \tilde{\xi}_4, \tilde{\xi}_5, \tilde{\xi}_6) = 0\) is asymptotically stable.

\section{State-Feedback Control Algorithm for Stabilization of Yaw Angle and Altitude}

Assume that the state vector \(\tilde{\xi}\) is available for measurement.

The control law \((u_1, u_2)\) can be chosen based on the cascade model using the feedback linearization method:
\begin{equation}\label{eq:u12}
\begin{bmatrix} u_1 \\ u_2 \end{bmatrix}
= b_1(\psi, \theta)^{-1} \left[ -q_1(\psi, \theta) - K_1\tilde{\xi}_1 - K_2\tilde{\xi}_2 \right],
\end{equation}
where the matrices \(K_1 = \operatorname{diag}(k_{11}, k_{12}) > 0\), \(K_2 = \operatorname{diag}(k_{21}, k_{22}) > 0\) are chosen based on specified performance criteria.

In this case, for the closed-loop system we have
\[
\begin{bmatrix} \dot{\tilde{\xi}}_1 \\ \dot{\tilde{\xi}}_2 \end{bmatrix}
= \begin{bmatrix} 0 & I_2 \\ -K_1 & -K_2 \end{bmatrix}
  \begin{bmatrix} \tilde{\xi}_1 \\ \tilde{\xi}_2 \end{bmatrix}
= F \begin{bmatrix} \tilde{\xi}_1 \\ \tilde{\xi}_2 \end{bmatrix},
\]
where \(I_2 = \begin{bmatrix} 1 & 0 \\ 0 & 1 \end{bmatrix}\) denotes the \(2 \times 2\) identity matrix. Assigning the desired eigenvalues to the matrix \(F\) will unambiguously determine the regulator parameters \(K_1\) and \(K_2\). Regrouping the state variables yields a model of the closed-loop system that is more convenient for synthesis:
\begin{equation}\label{eq:blockdiag}
\begin{bmatrix} \dot{\tilde{\xi}}_{11} \\ \dot{\tilde{\xi}}_{21} \\ \dot{\tilde{\xi}}_{12} \\ \dot{\tilde{\xi}}_{22} \end{bmatrix}
= \begin{bmatrix}
    0 & 1 & 0 & 0 \\
    -k_{11} & -k_{21} & 0 & 0 \\
    0 & 0 & 0 & 1 \\
    0 & 0 & -k_{12} & -k_{22}
  \end{bmatrix}
  \begin{bmatrix} \tilde{\xi}_{11} \\ \tilde{\xi}_{21} \\ \tilde{\xi}_{12} \\ \tilde{\xi}_{22} \end{bmatrix}
= \begin{bmatrix} F_1 & 0 \\ 0 & F_2 \end{bmatrix}
  \begin{bmatrix} \tilde{\xi}_{11} \\ \tilde{\xi}_{21} \\ \tilde{\xi}_{12} \\ \tilde{\xi}_{22} \end{bmatrix},
\end{equation}
where, due to the block-diagonal structure of the state matrix, the transient performance indices in the altitude \(z\) and yaw angle \(\psi\) loops can be tuned independently by the appropriate choice of the elements of matrices \(K_1\) and \(K_2\).

Noteworthily, due to the known structure of matrices \(q_1\) and \(b_1\), the control law can be rewritten as
\begin{align*}
\begin{bmatrix} u_1 \\ u_2 \end{bmatrix}
&= \begin{bmatrix}
     \dfrac{1}{\cos\theta\cos\psi} & 0 \\
     0 & 1
   \end{bmatrix}
   \left( g(1-\cos\theta\cos\psi) - K_1\tilde{\xi}_1 - K_2\tilde{\xi}_2 \right) \\
&= \begin{bmatrix}
     g\left(\dfrac{1}{\cos\theta\cos\psi}-1\right) + \dfrac{1}{\cos\theta\cos\psi}\bigl(-k_{11}\tilde{\xi}_{11}-k_{21}\tilde{\xi}_{21}\bigr) \\
     -k_{12}\tilde{\xi}_{12} - k_{22}\tilde{\xi}_{22}
   \end{bmatrix}.
\end{align*}
The yaw angle loop is shown to be stabilized by an independent PD controller:
\begin{equation}\label{eq:u2}
u_2 = -k_{12}\tilde{\xi}_{12} - k_{22}\tilde{\xi}_{22}.
\end{equation}
The altitude loop requires imposing a constraint on the signal \(u_1\) ensuring condition \eqref{eq:beta}. Let the nominal altitude controller be a PD controller with saturation:
\begin{equation}\label{eq:u1sat}
u_1 = \operatorname{sat}_L \left( \frac{1}{\cos\theta\cos\psi} \left( g - k_{11}\tilde{\xi}_{11} - k_{21}\tilde{\xi}_{21} \right) - g \right),
\end{equation}
where the function \(\operatorname{sat}_L(\cdot)\) denotes saturation with level \(L\). For \(L = \alpha g\) with an auxiliary parameter \(\alpha < 1\), condition \eqref{eq:beta} is ensured:
\[
0 < 1 - \alpha \le \beta(t) \le 1 + \alpha.
\]

In this case, the dynamics of the closed-loop altitude system will have the form
\[
\begin{bmatrix} \dot{\tilde{\xi}}_{11} \\ \dot{\tilde{\xi}}_{21} \end{bmatrix}
= \begin{bmatrix} 0 & 1 \\ 0 & 0 \end{bmatrix}
  \begin{bmatrix} \tilde{\xi}_{11} \\ \tilde{\xi}_{21} \end{bmatrix}
+ \begin{bmatrix} 0 \\ 1 \end{bmatrix}
  \cos\theta\cos\psi\,
  \operatorname{sat}_L \left( \frac{1}{\cos\theta\cos\psi} \left( g - k_{11}\tilde{\xi}_{11} - k_{21}\tilde{\xi}_{21} \right) - g \right),
\]
from which local input-to-state stability can be shown:
\[
\ddot{z} + k_{21}\dot{z} + k_{11}z = g(1 - \cos\theta\cos\psi),
\]
where the input is a function of roll \(\theta\) and pitch \(\psi\), which equals zero at the equilibrium state \(\theta = 0\) and \(\psi = 0\).

However, this property holds only for those roll and pitch angles for which the argument of the saturation function in the control coincides with its output value:
\[
\operatorname{sat}_L \left( \frac{1}{\cos\theta\cos\psi} (g - k_{11}\tilde{\xi}_{11} - k_{21}\tilde{\xi}_{21}) - g \right)
= \frac{1}{\cos\theta\cos\psi} (g - k_{11}\tilde{\xi}_{11} - k_{21}\tilde{\xi}_{21}) - g.
\]
Then we can find a constraint on the roll and pitch angles, assuming \(\tilde{\xi}_{11} = 0\) and \(\tilde{\xi}_{21} = 0\) at equilibrium:
\[
-\alpha g \le \frac{1}{\cos\theta\cos\psi} g - g \le \alpha g,
\]
from which we obtain the necessary condition
\[
\frac{1}{1 - \alpha} \ge \cos\theta\cos\psi \ge \frac{1}{1 + \alpha}.
\]
Since \(\alpha < 1\), then \(\frac{1}{1 - \alpha} > 1\), and the left inequality can be neglected. Thus, we have the condition
\[
\cos\theta\cos\psi \ge \frac{1}{1 + \alpha},
\]
the physical meaning of which is that the bounded control action is sufficient to balance the forces arising from roll and pitch deviations in the equilibrium at the desired altitude.

\section{State-Feedback Control Algorithm for Stabilization of Robot Coordinates in the Horizontal Plane}

Choose the control law \((u_3, u_4)\) based on the cascade model using the feedback linearization method:
\begin{equation}\label{eq:u34}
\begin{bmatrix} u_3 \\ u_4 \end{bmatrix}
= b_{22}(\phi, \psi, \theta)^{-1} \left[ -q_2(\phi, \psi, \theta, \dot{\phi}, \dot{\psi}, \dot{\theta}) - b_{21}(\phi, \psi, \theta) u_2 - K_3\tilde{\xi}_3 - K_4\tilde{\xi}_4 - K_5\tilde{\xi}_5 - K_6\tilde{\xi}_6 \right],
\end{equation}
where the matrices \(K_3 = \operatorname{diag}(k_{31}, k_{32}) > 0\), \(K_4 = \operatorname{diag}(k_{41}, k_{42}) > 0\), \(K_5 = \operatorname{diag}(k_{51}, k_{52}) > 0\), \(K_6 = \operatorname{diag}(k_{61}, k_{62}) > 0\) are chosen based on specified performance criteria in a special manner (see below), and the control signal \(u_2\) is defined in \eqref{eq:u2}. For the closed-loop system we obtain
\begin{equation}\label{eq:closed34}
\begin{bmatrix} \dot{\tilde{\xi}}_3 \\ \dot{\tilde{\xi}}_4 \\ \dot{\tilde{\xi}}_5 \\ \dot{\tilde{\xi}}_6 \end{bmatrix}
= \begin{bmatrix}
    0 & I_2 & 0 & 0 \\
    0 & 0 & \beta(t)I_2 & 0 \\
    0 & 0 & 0 & I_2 \\
    -K_3 & -K_4 & -K_5 & -K_6
  \end{bmatrix}
  \begin{bmatrix} \tilde{\xi}_3 \\ \tilde{\xi}_4 \\ \tilde{\xi}_5 \\ \tilde{\xi}_6 \end{bmatrix}.
\end{equation}
Note that the state matrix in \eqref{eq:closed34} is nonstationary; this means that the question of choosing the feedback coefficients is nontrivial and requires separate consideration.

Rewrite \eqref{eq:closed34} by regrouping the state vector similarly to \eqref{eq:blockdiag}:
\begin{equation}\label{eq:closed34_grouped}
\begin{bmatrix}
\dot{\tilde{\xi}}_{31} \\ \dot{\tilde{\xi}}_{41} \\ \dot{\tilde{\xi}}_{51} \\ \dot{\tilde{\xi}}_{61} \\
\dot{\tilde{\xi}}_{32} \\ \dot{\tilde{\xi}}_{42} \\ \dot{\tilde{\xi}}_{52} \\ \dot{\tilde{\xi}}_{62}
\end{bmatrix}
=
\begin{bmatrix}
0 & 1 & 0 & 0 & 0 & 0 & 0 & 0 \\
0 & 0 & \beta(t) & 0 & 0 & 0 & 0 & 0 \\
0 & 0 & 0 & 1 & 0 & 0 & 0 & 0 \\
-k_{31} & -k_{41} & -k_{51} & -k_{61} & 0 & 0 & 0 & 0 \\
0 & 0 & 0 & 0 & 0 & 1 & 0 & 0 \\
0 & 0 & 0 & 0 & 0 & 0 & \beta(t) & 0 \\
0 & 0 & 0 & 0 & 0 & 0 & 0 & 1 \\
0 & 0 & 0 & 0 & -k_{32} & -k_{42} & -k_{52} & -k_{62}
\end{bmatrix}
\begin{bmatrix}
\tilde{\xi}_{31} \\ \tilde{\xi}_{41} \\ \tilde{\xi}_{51} \\ \tilde{\xi}_{61} \\
\tilde{\xi}_{32} \\ \tilde{\xi}_{42} \\ \tilde{\xi}_{52} \\ \tilde{\xi}_{62}
\end{bmatrix}.
\end{equation}
The state matrix \(F_{34} = \begin{bmatrix} F_3 & 0 \\ 0 & F_4 \end{bmatrix}\) is block-diagonal:
\[
F_3 =
\begin{bmatrix}
0 & 1 & 0 & 0 \\
0 & 0 & \beta(t) & 0 \\
0 & 0 & 0 & 1 \\
-k_{31} & -k_{41} & -k_{51} & -k_{61}
\end{bmatrix},\quad
F_4 =
\begin{bmatrix}
0 & 1 & 0 & 0 \\
0 & 0 & \beta(t) & 0 \\
0 & 0 & 0 & 1 \\
-k_{32} & -k_{42} & -k_{52} & -k_{62}
\end{bmatrix},
\]
and its Hurwitz property (asymptotic stability of the system with such a state matrix) is determined by the Hurwitz property of matrices \(F_3\) and \(F_4\).

Since, according to \eqref{eq:beta}, the parameter \(\beta(t)\) is bounded both from above and below by known positive numbers, we can choose the constant parameters \(k\) to ensure asymptotic stability of the zero equilibrium in the system \eqref{eq:closed34_grouped} or \eqref{eq:closed34}. A similar statement can be found in articles \cite{1,2}, that can be proved based on the proof of an extended version of the Dayawansa lemma \cite{3,4}. The closest result for computing the matrix \(K\) is presented in \cite{5} using the backstepping method. Next, we present the proof of a statement that defines a method for choosing stabilizing feedback.

\textbf{Statement 2.}
Consider an auxiliary system of the form
\begin{align*}
\dot{\chi}_1 &= \chi_2,\\
\dot{\chi}_2 &= \beta(t)\chi_3,\\
\dot{\chi}_3 &= \chi_4,\\
\dot{\chi}_4 &= u,
\end{align*}
where \(\chi = \operatorname{col}(\chi_1, \chi_2, \chi_3, \chi_4) \in \mathbb{R}^4\), and the variable coefficient \(\beta(t)\) satisfies the double inequality \eqref{eq:beta}. The vector of constant parameters \(k = \operatorname{col}(k_1, k_2, k_3, k_4)\) is such that the control law
\begin{equation}\label{eq:ustab}
u = k^T \chi
\end{equation}
ensures asymptotic stability of the equilibrium \(\chi = 0\) for any initial conditions \(\chi(0)\).

\textbf{Proof of Statement 2.}
Consider a change of variables following the backstepping method:
\begin{align*}
y_1 &= \chi_1,\\
y_2 &= \chi_2 + \alpha_1 \chi_1,\\
y_3 &= \chi_3 + \alpha_2 y_2,\\
y_4 &= \chi_4 + \alpha_3 y_3.
\end{align*}
Then in the new basis \((y_1, y_2, y_3, y_4)\) the system takes the form
\begin{align*}
\dot{y}_1 &= -\alpha_1 y_1 + y_2,\\
\dot{y}_2 &= -\alpha_1^2 y_1 + (\alpha_1 - \alpha_2\beta) y_2 + \beta y_3,\\
\dot{y}_3 &= -\alpha_1^2\alpha_2 y_1 + \alpha_2(\alpha_1 - \alpha_2\beta) y_2 + (\alpha_2\beta - \alpha_3) y_3 + y_4,\\
\dot{y}_4 &= -\alpha_1^2\alpha_2\alpha_3 y_1 + \alpha_2\alpha_3(\alpha_1 - \alpha_2\beta) y_2 + \alpha_3(\alpha_2\beta - \alpha_3) y_3 + \alpha_3 y_4 + u.
\end{align*}
We choose the control law \(u = -\alpha_4 y_4\), which corresponds to formula \eqref{eq:ustab} where
\[
k_1 = -\alpha_1\alpha_2\alpha_3\alpha_4,\quad
k_2 = -\alpha_2\alpha_3\alpha_4,\quad
k_3 = -\alpha_3\alpha_4,\quad
k_4 = -\alpha_4.
\]
Then for the closed-loop system we obtain
\begin{equation}\label{eq:Phi}
\begin{bmatrix} \dot{y}_1 \\ \dot{y}_2 \\ \dot{y}_3 \\ \dot{y}_4 \end{bmatrix}
= \Phi \begin{bmatrix} y_1 \\ y_2 \\ y_3 \\ y_4 \end{bmatrix},
\end{equation}
with the state matrix \(\Phi\) having the form
\[
\Phi =
\begin{bmatrix}
-\alpha_1 & 1 & 0 & 0 \\
-\alpha_1^2 & \alpha_1 - \alpha_2\beta & \beta & 0 \\
-\alpha_1^2\alpha_2 & \alpha_2(\alpha_1 - \alpha_2\beta) & \alpha_2\beta - \alpha_3 & 1 \\
-\alpha_1^2\alpha_2\alpha_3 - \alpha_1\alpha_2\alpha_3\alpha_4 & \alpha_2\alpha_3(\alpha_1 - \alpha_2\beta) - \alpha_2\alpha_3\alpha_4 & \alpha_3(\alpha_2\beta - \alpha_3) - \alpha_3\alpha_4 & \alpha_3 - \alpha_4
\end{bmatrix}.
\]
The coefficients guaranteeing asymptotic stability of the zero equilibrium of system \eqref{eq:Phi} are chosen as follows.

\textbf{Step 1.} Choose an arbitrary positive number \(\alpha_1\).

\textbf{Step 2.} Choose the parameter \(\alpha_2\) based on input-to-state stability of the subsystem \((y_1, y_2)\). With zero input \(y_3\), the subsystem has the form
\begin{equation}\label{eq:subsys12}
\begin{bmatrix} \dot{y}_1 \\ \dot{y}_2 \end{bmatrix}
= \begin{bmatrix} -\alpha_1 & 1 \\ -\alpha_1^2 & \alpha_1 - \alpha_2\beta \end{bmatrix}
  \begin{bmatrix} y_1 \\ y_2 \end{bmatrix}.
\end{equation}

\textbf{Lemma 1.}
A positive number \(\alpha_2^*\) exists such that for all \(\alpha_2 > \alpha_2^*\) system \eqref{eq:subsys12} is asymptotically stable.

\textbf{Proof of Lemma 1.}
Consider the Lyapunov function \(V_2 = y_1^2 + y_2^2\). Differentiating \(V_2\), we obtain
\[
\dot{V}_2 = -\alpha_1 V_2 +
\begin{bmatrix} y_1 \\ y_2 \end{bmatrix}^T
\begin{bmatrix} -\alpha_1 & 1 - \alpha_1^2 \\ 1 - \alpha_1^2 & 3\alpha_1 - 2\alpha_2\beta \end{bmatrix}
\begin{bmatrix} y_1 \\ y_2 \end{bmatrix}.
\]
We choose the parameter \(\alpha_2\) from the condition of negative definiteness of the matrix
\[
\begin{bmatrix} -\alpha_1 & 1 - \alpha_1^2 \\ 1 - \alpha_1^2 & 3\alpha_1 - 2\alpha_2\beta \end{bmatrix}:
\]
\[
\alpha_2 > \frac{3\alpha_1^2 + (\alpha_1^2 - 1)^2}{2\alpha_1\beta}
      \ge \frac{3\alpha_1^2 + (\alpha_1^2 - 1)^2}{2\alpha_1\beta_{\max}} =: \alpha_2^*.
\]

\textbf{Step 3.} We choose the parameter \(\alpha_3 > \alpha_3^*\) based on input-to-state stability of the subsystem \((y_1, y_2, y_3)\) with zero input \(y_4\), by reapplying the arguments of the lemma to the Lyapunov function \(V_3 = y_1^2 + y_2^2 + y_3^2 = V_2 + y_3^2\).

\textbf{Step 4.} Similarly to steps 3 and 4, we choose the parameter \(\alpha_4 > \alpha_4^*\).

Thus, Statement 2 is proved.

Based on Statement 2, we can choose the coefficients \(K_3, K_4, K_5, K_6\) ensuring the stability of a system of the form \eqref{eq:closed34_grouped} and, as a consequence, asymptotic stability of \eqref{eq:closed34}.

\textbf{Remark 1.}
Noteworthily, the choice of feedback coefficients is complicated by the need to account for inequalities associated with the nonstationary coefficient \(\beta(t)\). Moreover, synthesis of a robust output-feedback control algorithm requires construction of a state observer, whose tuning to ensure closed-loop stability would require even greater effort.

\section{Dynamic State-Feedback Control Algorithm for Stabilization of Robot Coordinates in Space}

\textbf{Step 1.} We introduce new variables
\[
\zeta_1 = \begin{bmatrix} \tilde{\xi}_1 \\ \tilde{\xi}_3 \end{bmatrix},\quad
\zeta_2 = \begin{bmatrix} \tilde{\xi}_2 \\ \tilde{\xi}_4 \end{bmatrix},\quad
\zeta_3 = \begin{bmatrix} \tilde{\xi}_3 \\ \tilde{\xi}_5 \end{bmatrix},\quad
\zeta_4 = \begin{bmatrix} \tilde{\xi}_4 \\ \tilde{\xi}_6 \end{bmatrix},
\]
then the complete quadcopter motion model takes the form
\begin{align}
\dot{\zeta}_1 &= \zeta_2, \\
\dot{\zeta}_2 &= \begin{bmatrix} a_1(\psi,\theta) \\ 0 \end{bmatrix}
               + \begin{bmatrix} b_1(\psi,\theta) & 0 \\ 0 & \beta I_2 \end{bmatrix}
                 \begin{bmatrix} u_1 \\ u_2 \end{bmatrix}, \\
\dot{\zeta}_3 &= \zeta_4, \\
\dot{\zeta}_4 &= q_2(\phi,\psi,\theta,\dot{\phi},\dot{\psi},\dot{\theta}) + b_{21}(\phi,\psi,\theta)u_2 + b_{22}(\phi,\psi,\theta)
                \begin{bmatrix} u_3 \\ u_4 \end{bmatrix}.
\end{align}

\textbf{Step 2.} Let the control input vector \(\begin{bmatrix} u_1 \\ u_2 \end{bmatrix}\) be the output of two integrators with some input \(v_{12} = \begin{bmatrix} v_1 \\ v_2 \end{bmatrix}\) to be chosen later:
\[
\dot{u}_{12} = \rho_{12},\quad \dot{\rho}_{12} = v_{12}.
\]

\textbf{Step 3.} The quadcopter motion model can be represented in normal form.

\textbf{Statement 3.}
The aggregated quadcopter motion model can be represented in normal form:
\begin{align}
\dot{\zeta}_1 &= \zeta_2, \\
\dot{\zeta}_2 &= \zeta_3, \\
\dot{\zeta}_3 &= \zeta_4, \\
\dot{\zeta}_4 &= q_4(\zeta, \phi) + b_4(\zeta, \phi)U,
\end{align}
where \(\zeta = \operatorname{col}(\zeta_1, \zeta_2, \zeta_3, \zeta_4) \in \mathbb{R}^{16}\) is the state vector, \(U = \operatorname{col}(v_1, v_2, u_3, u_4)\) is the control input vector, and the functions \(q_4(\zeta, \phi)\) and \(b_4(\zeta, \phi)\) possess the following properties:
\[
q_4(0, \phi) = 0,
\]
the matrix \(b_4(\zeta, \phi)\) is nonsingular for all values of its arguments, and
\[
b_4(0, \phi) = g \begin{bmatrix} 0 & 0 \\ 0 & 0 \\ 1 & 0 \\ 0 & 1 \end{bmatrix} \beta \begin{bmatrix} -\cos\phi & -\sin\phi \\ -\sin\phi & \cos\phi \end{bmatrix}.
\]

\textbf{Proof of Statement 3.} can be obtained by sequentially computing the derivatives of the variables \(\zeta_3 = \dot{\zeta}_2\) and \(\zeta_4 = \dot{\zeta}_3\).

The final control law will have the form:
\begin{align}
\begin{bmatrix} F_1 \\ F_2 \\ F_3 \\ F_4 \end{bmatrix}
&= \frac{1}{4}
   \begin{bmatrix}
     1 & 1 & -1 & -1 \\
     1 & -1 & 1 & -1 \\
     1 & 1 & 1 & 1 \\
     -1 & -1 & -1 & 1
   \end{bmatrix}
   \begin{bmatrix}
     m & 0 & 0 & 0 \\
     0 & \dfrac{J_{\phi}}{C} & 0 & 0 \\
     0 & 0 & \dfrac{J_{\psi}}{\ell} & 0 \\
     0 & 0 & 0 & J_{\theta}
   \end{bmatrix}
   \begin{bmatrix} u_1 + g \\ u_2 \\ u_3 \\ u_4 \end{bmatrix}, \label{eq:F_final}\\
\begin{bmatrix} v_1 \\ v_2 \\ u_3 \\ u_4 \end{bmatrix}
&= b_4(\zeta, \phi)^{-1} \left[ -q_4(\zeta, \phi) - \gamma_1\zeta_1 - \gamma_2\zeta_2 - \gamma_3\zeta_3 - \gamma_4\zeta_4 \right], \label{eq:U_final}
\end{align}
where the choice of parameters \(\gamma_1, \gamma_2, \gamma_3, \gamma_4 > 0\) is significantly simplified, since it is determined by the specified performance criteria of the closed-loop system, whose model has the form:
\begin{equation}\label{eq:closed_zeta}
\begin{bmatrix} \dot{\zeta}_1 \\ \dot{\zeta}_2 \\ \dot{\zeta}_3 \\ \dot{\zeta}_4 \end{bmatrix}
= \begin{bmatrix}
    0 & I_4 & 0 & 0 \\
    0 & 0 & I_4 & 0 \\
    0 & 0 & 0 & I_4 \\
    -\gamma_1 I_4 & -\gamma_2 I_4 & -\gamma_3 I_4 & -\gamma_4 I_4
  \end{bmatrix}
  \begin{bmatrix} \zeta_1 \\ \zeta_2 \\ \zeta_3 \\ \zeta_4 \end{bmatrix},
\end{equation}
from which it becomes obvious that the parameters \(\gamma_1, \gamma_2, \gamma_3, \gamma_4\) can be chosen as the corresponding coefficients of standard Butterworth or Newton characteristic polynomials, or computed using the modal control method.

\section{Conclusion}

This study addresses the problem of dynamic point positioning of a quadcopter using two algorithms. The first one is based on the complete quadcopter motion model that accounts for the dynamics of all linear and angular coordinates, which generalizes the result of \cite{1} to the case of a varying yaw angle. The second one, a dynamic control algorithm, is characterized by a simplified regulator tuning methodology. The simplification stems from the fact that, in the synthesis of a robust control algorithm similar to \cite{1,2}, the structure of the state observer does not contain any nonstationary gains, which significantly facilitates the choice of its parameters.

\end{document}